\documentclass[aps,pra,showpacs,floats,twocolumn,unsortedaddress,floatfix,footinbib]{revtex4-1}
\usepackage{amsmath,amsfonts,amssymb,amsthm,stmaryrd,bm}
\usepackage{mathrsfs}
\usepackage{mathtools,wasysym}
\usepackage{accents}
\usepackage{graphicx}
\usepackage{hyperref}
\usepackage{breakurl}
\usepackage{color}
\usepackage{tikz-cd}

\newcounter{dummy}
\usepackage{enumitem}
\makeatletter
\newcommand\myitem[1][]{\item[#1]\refstepcounter{dummy}\def\@currentlabel{#1}}
\makeatother

\theoremstyle{plain}

\newtheorem*{lem*}{Lemma}
\newtheorem*{cor*}{Corollary}
\newtheorem*{thm*}{Theorem}
\newtheorem{prop}{Proposition}

\theoremstyle{definition}

\theoremstyle{remark}

\let\origmaketitle\maketitle
\def\maketitle{
  \begingroup
  \def\uppercasenonmath##1{} 
  \let\MakeUppercase\relax 
  \origmaketitle
  \endgroup
}

\newcommand{\defeq}{ {\kern 0.2em}:{\kern -0.5em}={\kern 0.2 em} }  
\newcommand{\eqdef}{ {\kern 0.2em}={\kern -0.5em}:{\kern 0.2 em} }  
\newcommand{\setof}[2]{\left\{ #1 \;\middle|\; #2 \right\}}  
\newcommand{\bigO}{{\mathcal O}}

\newcommand{\cc}[1]{\overline{#1}}  

\newcommand{\Arr}[3]{{#2}\colon {#1} \rightarrow {#3}} 

\newcommand{\Reals}{{\mathbb R}}    
\newcommand{\Cmplxs}{{\mathbb C}}   

\newcommand{\subdiff}
{\underline{D}}
\newcommand{\supdiff}
{\overline{D}}
\newcommand{\pair}[2]{\left\langle {#1} \,,\, {#2} \right\rangle}
\DeclareMathOperator{\re}{\mathrm{Re}}

\newcommand{\inpr}[2]{\left\langle {#1} \middle| {#2} \right\rangle}    
\newcommand{\Dbraket}[3]{\left\langle {#1} \middle| {#2} \middle| {#3} \right\rangle}

\DeclareMathOperator{\Spec}{\mathrm{spec}}

\newcommand{\tgt}[1]{{#1}^{\scriptscriptstyle\odot}}

\newcommand{\wh}[1]{ \widehat{#1} }
\newcommand{\hist}{\mathsf{Hist}}
\newcommand{\Str}{\mathsf{Str}}
\newcommand{\xs}{{\Delta}}
\newcommand{\zero}{{\mathscr Z}}

\newcommand{\slct}[1]
{\left[ {#1} \right] }
\newcommand{\FHK}{F^{\text{HK}}}

\newcommand{\V}
{B'}
\newcommand{\VC}
{B_\Cmplxs'}
\newcommand{\Vzero}{{\mathcal V}_0}

\begin{document}

\title{
  A bivariate view of Kohn-Sham iteration 
  and the case for potential mixing
}
\author{Paul~E.~Lammert}
\email{lammert@psu.edu}
\affiliation{Department of Physics, 104B Davey Lab \\ 
Pennsylvania State University \\ University Park, PA 16802-6300}
\date{July 4, 2020}
\date{\today}
\begin{abstract}
A bivariate perspective on Kohn-Sham density functional theory is proposed,
treating potential and density as simultaneous independent variables,
and used to make fruitful connection between Lieb's rigorous foundational
framework and practical Kohn-Sham computation.
Support is found for potential-mixing schemes, but not for 
more standard density-mixing. Under presumably-generic
conditions, total energy can be lowered from one iteration to the next.
Density, intrinsic and total energy are analytic functions of
the noninteracting potential on the open set of potentials having
a single isolated ground spin-multiplet.
\end{abstract}
\maketitle

\section{Introduction}
Over the past half-century, ground-state density functional theory (DFT) in the
dominant Kohn-Sham\cite{Kohn+Sham} (KS) form has 
developed into a ubiquitous tool in physics, chemistry, materials science, and 
beyond\cite{Hohenberg+Kohn,Parr+Yang,Dreizler+Gross,Koch+Holthausen,Capelle06,Burke-12}.
Distinguishing characteristics of the KS formulation are (i) 
at the theoretical level, a splitting of the intrinsic energy functional into noninteracting
and Hartree-exchange-correlation functionals, and (ii) at the computational level,
a particular class of iterative procedures connected to the implicit and explicit
forms, respectively, of those functionals.
However, as distinct from the problem of accelerating convergence, the simple fundamental
question of why one should expect the iterative scheme to progress toward a solution
has received scant attention.

This Letter presents a deeper understanding of practical KS iteration and its relation to
the rigorous foundational program initiated by Lieb\cite{Lieb83}, while maintaining a
focus on what is in practice computable.
Simply giving densities and potentials equitable status as independent variables
--- the bivariate perspective --- is instrumental to the project.
This is analogous to the independence of position and momentum in Hamiltonian mechanics,
even though they are not independent along physical trajectories.
Despite the even-handed approach, strong grounds are found for preferring potential
over density for purposes of guiding iteration, and potential-mixing strategies over
density-mixing strategies.
No clear rationale for density-mixing is discerned.
On the other hand, within a miminal abstract framework, an inequality
is derived (see Prop. \ref{prop:progress}) suggesting that potential-mixing can
make progress in the ordinary energetic sense under presumably-generic conditions
[see Eq. \ref{eq:excess-energy})],
and in a practically verifiable way. More concretely --- i.e., bringing the full machinery of
quantum mechanics to bear --- it is shown (Prop. \ref{prop:progress-Lieb}) not only that
this expectation is borne out, but that all functions relevant at the density-functional
level are {\em analytic} as functions of potential, 
where the noninteracting ground manifold is a single isolated spin-multiplet.
Sophisticated convergence acceleration
algorithms\cite{Pulay-80,Kudin+02,Walker+Ni-11,Garza+Scuseria-12} have
been developed and implemented in software packages.
They are based on {\em assumptions} of smoothness and fundamental
soundness of the algorithm being accelerated; it is precisely the latter that
are under investigation here. The present results provide some justification
for those assumptions, but not for the most common density-mixing schemes.

General DFT has its proper limited language, which makes no reference to quantum mechanics.
Much of the discussion here (before Prop. \ref{prop:progress-Lieb})
is at that level, in a minimally axiomatized abstract framework.
The primary interpretation of interest is the standard ``$L^1\cap L^3$'' interpretation of
Lieb\cite{Lieb83,Eschrig,vanLeeuwen03,Lammert10a}
(see Appendix \ref{sec:app-Lieb} for a more information).
for a fixed, finite number of particles.
However, alternative interpretations are also of interest, including,
infinitely many particles in a periodic potential, variants in bounded domains,
discrete space versions, appropriate forms of nonzero-temperature quantum DFT\cite{Gonis18},
classical density functional theory\cite{Lischner+Arias08},
Kohn-Sham theory for fractional quantum Hall effect\cite{Hu+Jain-19},
and perhaps some where $\rho$ and $v$ are read as something entirely
different from density and potential.
Laestadius {\it et al.}\cite{Laestadius+18-JCP} have recently proposed an alternative
abstract approach, based on a Moreau-Yosida regularization of DFT\cite{Kvaal+14},
but with aims somewhat different from here.

{\it Background ---}
In general DFT, the basic quantum mechanical $N$-body ground state problem is phrased as follows.
Let $F(\rho)$ be the minimum kinetic-plus-interaction energy over all the states of the
$N$ particles having density function $\rho(x)$.
Then, with $\tgt{v}$ an external one-body potential of interest,
minimize $F(\rho) + \int \tgt{v}\rho\, dx$ over $\rho$.
The minimum value is the ground energy $E(\tgt{v})$,
any minimizing $\rho$ is a ground state density, and is characterized
by the Euler criterion, $DF(\rho) + \tgt{v} = 0$ ($D$ denotes a functional
derivative).
From the functional $F$ will flow the ground state energies and densities of all
the external potentials that may interest us.
Alas, this vision faces a number of difficulties, the most severe being lack
of an effective, direct, way to calculate $F(\rho)$ for even a single $\rho$.
Kohn-Sham theory attempts to circumvent this by splitting $F$ as $F_0 + \Phi$,
where $F_0$ is the counterpart of $F$ for a {\em noninteracting} system, and
$\Phi$ is known as Hartree-exchange-correlation energy.
$F_0$ is not {\em directly} accessible any more than is $F$,
but finding ground states for noninteracting particles in a given external potential $v$
is feasible. Information about $F_0$ can thus be obtained.
With it, and using some explicit approximation to $\Phi$, one might seek
to satisfy the Euler criterion for $F$ using $DF = DF_0 + D\Phi$.
A potential obstacle is the fact that there is no nontrivial topology not defined in
terms of the functionals $F_0$ and $F$ themselves, relative to which they have been
shown to be continuous. Worse, they are {\em nowhere continuous},
merely lower semicontinuous, functions on the Banach space $L^1(\Reals^3)\cap L^3(\Reals^3)$.
Recall that for a function $f$ on a normed space, $f$ being {\em lower semicontinuous}
(lsc) at $x$ means that whenever $x_n\to x$, then
$\lim_{n\to\infty} f(x_n) \ge f(x)$. (Upper semicontinuous, usc, has the opposite
inequality.)

Derivatives are thus expected to present some problems.
Suppose $f\colon V \rightarrow \Reals$ is a function on a normed vector space $V$.
A derivative of $f$ at $a$ is an approximation by a continuous affine function.
For instance, a linear functional $\lambda$ in the dual space $V'$ of
continuous linear functionals is the G-derivative $Df(a)$ if
$f(a+sx) = f(a) + s\pair{\lambda}{x} + o(s)$ for each $x\in V$.
Here, $\pair{\lambda}{x}$ denotes the natural pairing between vectors and dual vectors
(e.g., $\pair{v}{\rho} = \int v(x)\rho(x)\, dx$).
If $f$ is not regular enough at $a$ for that, a {\em unilateral} approximation
from below --- $f(x) \ge f(a) + s\pair{\lambda}{x} + o(s)$ --- may still be available.
Then, $\lambda$ is a G-{\it subgradient}\cite{Schirotzek,Penot}
({\it supergradient} for the opposite inequality). The $G$-{\it subdifferential},
$\subdiff f(a)$ (warning: not usual notation) is the set of all subgradients,
of which there may be many, or none.
For example, the absolute value $x\mapsto |x|$ has subdifferential $[-1,1]$ at zero.
Absolute value is convex, which recall, means all secant lines are on or above the graph.
For a convex function, and these have a special role here just as in thermodynamics,
$\lambda \in \subdiff f(a)$ implies that $f(a+x) \ge f(a) + \pair{\lambda}{x}$,
i.e., a global unilateral bound, not just an approximate asymptotic one.
In a DFT context, subdifferentials seem to be the best kind of affine approximation
available. Fortunately, it is in the nature of minimization-type problems
that subdifferentials suffice.
It also bears worth noting that the superdifferential of the energy corresponds to
ground densities. Accomodating the fact that the latter are not always unique
forces the use of unilateral, rather than ordinary bilateral, derivatives.

{\it Abstract framework ---}
The abstract framework within which we will work is layed out in the
following postulates/axioms {\textbf{\textsf{A}}} and {\textbf{\textsf{B}}}.
\begin{enumerate}
\item[{\textbf{\textsf{A}}}.] 
$B$ is a Banach (complete normed) space;
\begin{equation}
F_0,\Phi \colon B \xrightarrow{\text{lsc}} {\Reals}\cup\{\infty\}
\end{equation}
are lower semicontinuous (lsc); and $F_0$ and 
\begin{equation}
F = F_0 + \Phi
\end{equation}
are {\em convex}.
\end{enumerate}
Remarks.
For $B = L^1(\Reals^3)\cap L^3(\Reals^3)$ and fermions interacting via Coulomb
repulsion (Lieb interpretation),
the noninteracting, $F_0$, and interacting, $F$, intrinsic energy functionals
for mixed states are convex and lower semicontinuous.
In that context, $\Phi$ is Hartree-exchange-correlation energy.
We {\em cannot} assume that $\Phi$ is continuous if GGA\cite{Perdew+96,Tao+03}
exchange-correlation functionals are to be accomodated.
For our purposes, the interpretation of interest is not necessarily the exact theory.
Insofar as computational behavior is at stake, $\Phi$ should be taken to be
the implemented approximation.

Define the {\it ground energy} $E$ on the dual space $B'$ of $B$ via
\begin{equation}
E(v) = \inf_\rho \{ F(\rho) + \pair{v}{\rho} \}.  
\label{eq:E from F}
\end{equation}
$E$ is upper semicontinuous and concave (i.e., $-E$ is lsc and convex),
so the {\it superdifferential} $\supdiff E$, giving unilateral affine
approximation from above, will be relevant, instead of subdifferential.
It is convenient to package $F$ and $E$ together into the {\it excess energy}
\begin{equation}
\label{eq:excess-energy}
\xs(v,\rho) \defeq F(\rho) + \pair{v}{\rho} - E(v) \ge 0
\end{equation}
function on $B'\times B$, which is separately convex in its two arguments and jointly lsc.
$\xs(v,\rho)$ embodies the bivariate spirit, answering the question, ``how close to
the ground energy $E(v)$ can one get with states of density $\rho$?''
The zero set $\zero = \{\xs=0\}\subset B'\times B$ contains all possible solutions
of all possible ground density problems. If $(v,\rho)$ is in $\zero$,
we call it a {\it ground pair}.
$E_0$, $\xs_0$, $\zero_0$ are defined from $F_0$ in exactly the same way as
$E$, $\xs$, $\zero$ are defined from $F$.
In distinguishing between the two, we prefer the designations
`reference/perturbed' over `noninteracting/interacting'.

Since we wish to discuss iteration strategies from an abstract perspective,
a semi-formalized concept of practical computability, to be called {\it feasibility},
will be helpful. With the exception of unbounded search, ordinary computing elements,
arithmetic, loops, branching, etc., are to be considered feasible. A composition of
finitely-many feasible operations is feasible. To be considere feasible,
a partially-defined function should give, in bounded time, notification that
an out-of-domain argument is so.
In addition, guided by actual practice,
we simply postulate that certain context-specific functions
are feasible. Partly this is due to the fact that vectors in an infinite-dimensional
space, and real numbers for that matter, are not exactly representable in a computer.
\begin{enumerate}
\item[{\textbf{\textsf{B}}}.] Computations approximate points in $B$, $B'$ in norm.
  The following operations are feasible:
  vector addition, scalar multiplication and the pairing
  $\pair{\phantom{v}}{\phantom{\rho}}$ of $B'$ and $B$;
  also $E_0$, $[\supdiff E_0]_1$, $\Phi$, and $[\subdiff \Phi]_1$. 
\end{enumerate}
The final clause embodies the raw operations of KS computation.
$\supdiff E_0(v)$ is the {\em set} of ground densities for $v$ in the reference system.
But we might not want to insist that the computation provide them all.
The (unspecified and conceivably nondeterministic) selection operator $[\phantom{E_0}]_1$
delivers one sub- or super-gradient, if there are any.
The HXC energy $\Phi$, even $\subdiff \Phi$, is usually given by an explicit formula.
This list is important. If $\subdiff F$ were postulated to be feasible, an ordinary
gradient-descent algorithm would be a reasonable proposal.
The idea is that a proposed computational strategy ought to come with
a {\em warrant} that it is feasible. No means have been specified to
show that anything is {\em not} feasible (though strong suspicions might
well be in order).

\begin{figure}
  \centering
\includegraphics[width=80mm]{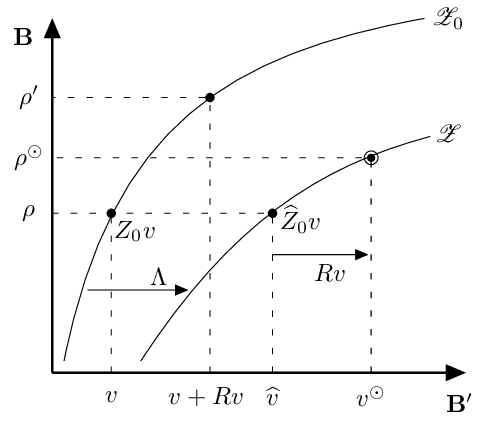}
\caption{
  Schematic representation of the bivariate perspective in the product space $B'\times B$.
  The zero excess energy sets $\zero_0$ and $\zero$ are indicated, along with some of the
  functions listed in Table \ref{tab:feasible}.
  The picture is, of course, not faithful in all aspects:
  $B$ and $B'$ are generally infinite-dimensional,
  $\zero$ and $\zero_0$ are not likely to be smooth, or even
  (single-valued) functions.
\label{fig:1}}
\end{figure}

{\it Walking on the ground pairs ---}
Here is the $\tgt{v}$-Problem:
given $\tgt{v}$ in $B'$, find $(\tgt{v},\tgt{\rho})$ in $\zero$, or a 
near enough approximation thereof (in norm sense, or in total energy sense
as discussed below).
The situation in the product space $B'\times B$ is schematically
illustrated in Fig. \ref{fig:1}. The curves represent the sets
$\zero_0$ and $\zero$ of reference and perturbed ground pairs.
Our ultimate interest is in $\zero$, but $\zero_0$ is more immediately
accesible, and plays an intermediary role. 
Table \ref{tab:feasible} gives a collection of feasible (partial) functions,
most of which are illustrated on Fig. \ref{fig:1}.
\begin{table}[h]
  \caption{Basic feasible functions/operations, described in the text.
    $\circ$ is the composition operator, $\pi_{B'}$ extracts $B'$ component,
    and $\rightharpoonup$ indicates a partial (not everywhere defined) function.
    \label{tab:feasible}
    }
  \centering
\begin{tabular}{lcl}
  \hline
  name & definition & type \\ \hline
  $Z_0$  & $v \mapsto (v, \slct{\supdiff E_0(v)} )$ &
 $B' \rightharpoonup \zero_0$ 
 \\  
   $\Lambda$ & $(v,\rho) \mapsto (v-\slct{\subdiff\Phi(\rho)}, \rho)$ &
 $\zero_0 \rightharpoonup \zero$
   \\
 $\wh{Z}_0$ & $\Lambda \circ Z_0$ &
 $B' \rightharpoonup \zero$
    \\
$(\wh{\phantom{v}})$ & $\pi_{B'}\circ \wh{Z}_0 $ &
 $ B' \rightharpoonup B'$
  \\
$R$  
& $\tgt{v} - (\wh{\phantom{v}})$  &
 $ B' \rightharpoonup B'$
  \\
  $\FHK$
  & $(v,\rho) \mapsto F(\rho)$
  & $\zero_0 \rightarrow \Reals$
\\ \hline
\end{tabular}
\end{table}

These contain a repackaging of the basic feasible operations postulated in
\textbf{\textsf{B}}.
$Z_0$, pairing a potential with a ground density for the reference system,
is a trivial rephrasing of $\slct{\supdiff E_0}$. 
$\Lambda$ puts the feasibility of $\subdiff\Phi$ to work, and is more interesting.
Since $F = F_0 + \Phi$, 
$-v\in\subdiff F_0(\rho)$ implies that $-v+\subdiff\Phi(\rho) \in \subdiff F(\rho)$.
($\subdiff F_0+\subdiff\Phi \subseteq \subdiff F$; the reverse inclusion is
delicate, but not needed.)
Computation of points in $\zero_0$ is given, by assumption, whereas $\Lambda$ generates
points in $\zero$ from them via the Euler criterion. Convexity of $F$ is the guarantor
that they {\em really are} zeros of $\xs$.
Summing up: given $v\in B'$, a feasible operation gives us
$(v,\rho) = Z_0v \in\zero_0$ and a second yields $(\wh{v},\rho) = \wh{Z}_0 v \in \zero$.
The first step fails if $v$ cannot bind $N$ particles. Failure of the second would
be a sort of $V$-representability problem. The possibility of such exceptional conditions
is the price to be paid for avoiding possibly unrealistically restrictive assumptions.
Insofar as the interest is in analyzing normal circumstances, this is tolerable as
long as the epithet ``exceptional'' is deserved.
Every time a reference problem is solved via $Z$, solution to a perturbed problem
is also made available --- $(\wh{v},\rho) \in \zero$.
This suggests a change of perspective on KS iterative computations:
Rather than viewing it as a sequence of approximate solutions to the given $\tgt{v}$-Problem,
we view it as a sequence of solutions to {\em approximate problems}. Points
on $\zero$ are generated in a peculiar manner and the task is to steer
the sequence so that the $B'$ components approach $\tgt{v}$.
The {\it residual} $Rv=\tgt{v}-\wh{v}$ is a kind of measure of proximity to solution:
if $(\wh{v},\rho) = (\tgt{v}-Rv,\rho)\in\zero$ then $Rv$ is the amount by which
$\tgt{v}$ must be perturbed to render $\rho$ a ground density.

The only function in the table which uses the postulated feasibility of
either $\Phi$ or $E_0$, as opposed to their subdifferentials, is $F^{\text{HK}}$,
which uses both. If $(v,\rho) \in \zero_0$, then
$0 = \xs_0(v,\rho) = F_0(\rho)+\pair{v}{\rho}-E_0(v)$, and therefore
\begin{equation}
F(\rho) = F^{\text{HK}}(v,\rho) = E_0(v)-\pair{v}{\rho}+\Phi(\rho).
\end{equation}
The superscript `HK' stands for `Hohenberg-Kohn', because this is closer to
the original\cite{Hohenberg+Kohn} intrinsic energy definition than the later
constrained-search formulation\cite{Levy79,Levy82}.
Note that, to obtain $F(\rho)$, auxiliary data consisting of a potential
partner in the {reference} system is needed.
There is no apparent feasible route from
$(v,\rho)$ in $\zero$ to $F(\rho)$, because $E(v)$ would be needed.
$\FHK$ can be used to monitor energetic {\em progress}.
Given $(v,\rho),(v',\rho')\in\zero_0$ a feasible test is available to
determine which of $\xs(\tgt{v},\rho)$ and $\xs(\tgt{v},\rho')$ is smaller,
namely,
$\xs(\tgt{v},\rho) - \xs(\tgt{v},\rho')
= \FHK(v,\rho)-\FHK(v',\rho')+\pair{\tgt{v}}{\rho-\rho'}$.
It is desirable to clarify the relation between this energetic idea of proximity to
a solution ot the one based on the residual which was introduced earlier.
ion. The following Proposition depends on lower semicontinuity,
and is proved in Appendix \ref{app:approx}.
\begin{prop}
  \label{prop:energetic-vs-residual}
If $\xs(\tgt{v},\rho) < \epsilon$, then $(v',\rho') \in \zero$ for
some $(v',\rho')$ satisfying $\|\tgt{v}-v'\|, \|\rho-\rho'\| < \sqrt{\epsilon}$.
Going in the other direction, if $E$ is locally Lipschitz continuous
then $(\wh{v},\rho) \in \zero$ and $\wh{v}\in U$ 
imply that \hbox{$\xs(\tgt{v},\rho) < (L+\|\rho\|) \| Rv\|$}
(Lipschitz constant $L$ on neighborhood $U$ of $\tgt{v}$). 
\end{prop}
Local Lipschitz continuity means that
there is a neighborhood $U$ of $\tgt{v}$ such that
$|E(v)-E(v')| < L \|v-v'\|$ whenever $v$ and $v'$ are in $U(\tgt{v})$.
Sufficient conditions for this are:
$E(v)$ is finite for every $v$ in $B'$, and
for some $M, c > 0$,  $F(\rho) > c\|\rho\|$ when $\|\rho\| > M$.
These have been established\cite{Lieb83} for the Lieb theory.
A moral of the Proposition is that a stopping (``convergence'') criterion based on the
apparently-infeasible excess energy is essentially nearly equivalent to 
a feasible one based on the residual.
If the cycle-to-cycle change in the input potential is closely related to $Rv$,
this provides some support to stopping criteria based on such changes by showing
that they actually have a disguised absolute character.

{\it Strategies ---}
Consider now how to select potentials to feed to $Z_0$.
Given the {\it history} of input-output pairs
\begin{align}
\hist_n 
&=  (v_1;\wh{Z}_0 v_1),(v_2;\wh{Z}_0 v_2),\ldots,(v_n;\wh{Z}_0 v_n)
\label{eq:history}
\\
&=  (v_1;\wh{v_1},\rho_1),(v_2;\wh{v_2},\rho_2),\ldots,(v_n;\wh{v_n},\rho_n),
\nonumber
\end{align}
what are good {\it strategies} for choosing $v_{n+1}$?
The simplest and most obvious is the {\it basic strategy}
\begin{equation}
\Str_1(\hist_n)  = {v}_n + R{v_n}.
\label{eq:plain vanilla}
\end{equation}
Effectively, $\Str_1$ embodies the hypothesis that
$\wh{v}-v$ varies little with $v$, and uses the last stage of $\hist$ to calculate it. 
Alas, $\Str_1$ has a well-known tendency toward charge sloshing instability. 

The cure prescribed by standard practice\cite{Martin,Primer-DFT} is most easily described using an
augmented kind of history,
\begin{equation}
\hist_n^+ \defeq (\rho^{\text{in}}_1;v_1;\wh{v_1},\rho_1),
\ldots,(\rho^{\text{in}}_n;v_n;\wh{v_n},\rho_n).
\label{eq:history+}
\end{equation}
With $0 < \lambda \le 1$, $\rho^{\text{in}}_k$
serves to {parametrize} $v_k$ according to 
\begin{equation}
\Str_\lambda^{\rho}(\hist_n^+) = \tgt{v} - \slct{\subdiff\Phi(\rho^{\text{in}}_{n+1})},
\label{eq:rho-mixing-1}
\end{equation}
where 
\begin{equation}
\rho^{\text{in}}_{n+1} \defeq \lambda \rho_{n} + (1-\lambda)\rho^{\text{in}}_{n}.
\label{eq:rho-mixing-2}
\end{equation}
For $\lambda=1$, $\Str^{\rho}_\lambda$ reduces to the strategy $\Str_1$.
%
An alternative, potential-mixing, strategy is
\begin{equation}
  \Str_\lambda^v = \lambda\, \Str_1 + (1-\lambda)\, \Str_0,
\label{eq:v-mixing}
\end{equation}
where $\Str_0$ is the trivial (but extremely stable!) {repeat} strategy
$\Str_{0}(\hist_n) = v_n$.
$\Str^v_\lambda$ follows the advice of $\Str_1$, but cautiously,
taking only a small step in the suggested direction $Rv$.
$\Str_\lambda^{\rho}$ also interpolates between $\Str_0$ and $\Str_1$.
Indeed, the two strategies differ to the extent that $\subdiff\Phi$ is
nonlinear on the segment $[\rho_n^{\text{in}},\rho_n]$.
However, I suggest that the way it does so is indirect, seemingly unnatural,
and has no clear rationale.

{\it Progress ---}
Seeming naturality of a strategy is a virtue, but not the only or most important one.
Suppose that $Z_0v_0 = (v_0,\rho_0)$ is in hand.
Define also $v_1 = v_0 + Rv_0$, and $\rho_1$ by $Z_0v_1 = (v_1,\rho_1)$.
The latter pair is simply the next member in the basic strategy $\Str_1$ sequence.
We now ask,
is there a family of densities $\rho_\lambda$ interpolating between $\rho_0$ and $\rho_1$,
such that (barring straightforward failures such as $Z_0$ returning empty)
$\xs(\tgt{v},\rho_\lambda) < \xs(\tgt{v},\rho_0)$ is guaranteed for some $\lambda$?
Note that subscripts on $v$ and $\rho$ are being used differently than in the
prevous section. Since only one step is being considered and $\lambda$ is not
necessarily integral, no confusion should result.

One possibility is a linear interpolation in density:
\begin{equation}
  \label{eq:rho-tilde}
\widetilde{\rho}_\lambda =  (1-\lambda){\rho}_0 + \lambda {\rho}_1
\end{equation}
for $0\le\lambda$.
It can be shown that
\begin{equation}
\frac{d}{d\lambda}\xs(\tgt{v},\widetilde{\rho}_\lambda)\Big|_{\lambda=0} < 0,
\label{eq:progress try 1}
\end{equation}
whenever the derivative exists.
This result was given by Wagner {\it et al.}\cite{Wagner+13}
and later corrected/rigorized by Laestadius {\it et al.}\cite{Laestadius+18-JCP}.
It is proved in Appendix \ref{app:progress} as Prop. \ref{appprop:progress-bad},
demonstrating that the minimal abstract axiomatization is enough.
Unfortunately, there is a fatal flaw to (\ref{eq:progress try 1})
as a basis of a strategy.
To be able to use it in a non-blind way, we must be able to test the value of
$\xs(\tgt{v},\widetilde{\rho}_\lambda) - \xs(\tgt{v},\widetilde{\rho}_0)$.
As previously discussed, the only evident feasible way to do that is to obtain
$\widetilde{\rho}_\lambda$ as the second component of a point on $\zero_0$, which
means we need to know a potential having $\widetilde{\rho}_\lambda$ as a ground density.
And, that is not forthcoming.
The family (\ref{eq:rho-tilde}) might deserve the name ``density-mixing'' more than
the strategy $\Str_\lambda^\rho$, which is feasible, and
could be considered a non-linear form of potential-mixing in disguise.

A second attempt to find a method of feasibly making progress
involves interpolation of the potential according to:
\begin{equation}
v_\lambda = (1-\lambda) v_0 + \lambda v_1  = v_0 = \lambda Rv_0,
\end{equation}
together with $\rho_\lambda$ defined implicitly via 
\begin{equation}
({v_\lambda},\rho_\lambda) = Z_0 v_\lambda.
\end{equation}
These densities also interpolate between $\rho_0$ and $\rho_1$, but 
are feasible by construction.
A strengthening of the following result is proved in
Appendix \ref{app:progress} as Prop. \ref{appprop:progress-good}.
\begin{prop}
\label{prop:progress}
With the preceding notation, assuming $\rho_\lambda$ exists and $Rv \neq 0$, 
\begin{equation}
  \label{eq:Delta-xs-upper-bd}
\xs(\tgt{v},\rho_\lambda) - \xs(\tgt{v},\rho_0) 
<
\xs(\wh{v_0},\rho_\lambda)
-\frac{1}{\lambda} \xs_0({v_0},\rho_\lambda).
\end{equation}
\end{prop}
Recall that $\xs_0$ and $\xs$ are everywhere non-negative.
The remarkable, and encouraging, aspect of the inequality
(\ref{eq:Delta-xs-upper-bd}) is the extra factor $\lambda^{-1}$ in
the negative term. One might well expect both excess energies to
be asymptotically quadratic in $\lambda$, giving an initial linear
decrease of the right-hand side.

In the standard Lieb interpretation, on the set of potentials for which the
noninteracting system has an isolated spin-multiplet ground manifold, 
not only this expectation, but even analytic behavior, can be verified.
For proof of the following, see Appendix \ref{sec:app-Lieb}.
\begin{prop}
\label{prop:progress-Lieb}
Let ${\mathscr V}_0 \subset L^{3/2}(\Reals^3)+L^\infty(\Reals^3)$ be the set
of external potentials $v$ in the standard interpretation such that the corresponding
reference system Hamiltonian has an isolated ground state eigenvalue and the ground
state manifold comprises a single spin multiplet, so that
there is a ground density, denoted $\rho[v]$.
\newline
(A)
${\mathscr V}_0$ is open. $\rho[v]$, $F(\rho[v])$, and 
$\xs(\tgt{v},\rho[v])$ are analytic functions of $v\in{\mathscr V}_0$.
\newline
(B)
Suppose $v_0 \in {\mathscr V}_0$. For 
$\lambda$ in some neighborhood $U$ of zero, $v_\lambda \defeq v_0 + \lambda Rv_0$
is in ${\mathscr V}_0$ and $\rho_\lambda \defeq \rho[v_\lambda]$ is unambiguous.
Assuming $Rv_0\neq 0$,
either $\rho_\lambda = \rho_0$ for $\lambda\in U$ [possible only if the
ground state(s) are eigenstates of $Rv_0$],
or $\frac{d}{d\lambda}\xs(\tgt{v},\rho_\lambda) < 0$. 
\end{prop}
Note that, the exceptional circumstance recognized in part (B) (in square brackets)
is forbidden by the Hohenberg-Kohn theorem, which is at present proven for
locally square-integrable potentials\cite{Garrigue-18}.

The point about which Prop. \ref{prop:progress-Lieb} turns is,
together with Prop. \ref{prop:progress},
that analyticity in quantum mechanical perturbation theory lifts unproblematically to
the density functional level.
The significance is the support it gives to potential-mixing strategies.
Most simply, one may repeatedly halve $\lambda$ and test
$\xs(\tgt{v},\rho_\lambda) - \xs(\tgt{v},\rho_0)$ until a negative value is found.
The asserted analyticity supports much more sophisticated schemes.
Restricting attention to a finite-dimensional subspace in ${\mathscr V}_0$
and introducing coordinates, analyticity can be expressed in the elementary
form of convergent power series.
Thus, although radius of convergence is an unaddressed aspect, this supports
fitting a quadratic function of $\lambda$ to find a line minimum, or even
multidimensional acceleration schemes in potential space.

The standard interpretation is adequate for molecules, but not for solids, since
it accomodates neither an infinite number of particles nor periodic potentials in
extended space. Generic lack of a spectral gap (metals) is another characteristic
of extended systems. In light of the gap condition in Prop. \ref{prop:progress-Lieb},
the development of a rigorous DFT for truly extended systems now seems more
interesting and urgent.

{\it Conclusion ---}
A bivariate perspective on Kohn-Sham iteration has been proposed here and
shown to be useful in bridging the divide between a rigorous foundation and
practical Kohn-Sham computations.
It may also be useful in more heuristic settings, such as development of
convergence acceleration algorithms.
Rigorous results, both in a minimal abstract axiomatic setup and,
much more strongly, in the standard Lieb interpretation,
give support to potential-mixing schemes, showing how they can
make progress in an energetic sense.
Density, intrinsic energy and excess energy are all {\em analytic}
functions of noninteracting potential on the open set of such potentials
having an isolated ground-state spin multiplet.
This helps explain how practical computations can be insulated from
nonsmoothness of the intrinsic energy.

\newtheorem{appthm}{Theorem}
\newtheorem{appcor}[appthm]{Corollary}
\newtheorem{appprop}[appthm]{Proposition}
\newcommand{\Hilb}{{\mathscr H}}
\newcommand{\Four}[1]
{\widetilde{#1}}
\newcommand{\UU}{{\mathcal U}}
\newcommand{\Hol}{\mathrm{Hol}}
\appendix

\section{Useful identities}
\label{app:identities}

This Section collects some identities which are proven by straighforward
manipulation, starting from the definition of excess energy, which recall is
\begin{equation}
  \label{eq:xs}
\xs({v},\rho) = F(\rho) + \pair{{v}}{\rho} - E({v}).
\end{equation}
{1} -- {3} hold in either the reference system
(in which case subscripts $0$ should be attached) or the perturbed system.

\smallskip
\noindent {1}.
{Cross-difference identity:}
\begin{align}
\xs(v,\rho)  
- \xs(v,\rho')  
+ & \xs(v',\rho')   
- \xs(v',\rho)  
\nonumber \\
& = \pair{v - v'}{\rho - \rho'}. 
\label{eq:cross-difference}
\end{align}
To derive this, note that
each of $v$, $v'$, $\rho$ and $\rho'$ appears 
on the left-hand side of (\ref{eq:cross-difference}) 
as an argument of two $\xs$'s, one with a minus sign. 
Thus, substituting the definition (\ref{eq:xs}),
all the $F$'s and $E$'s cancel out. tallying up the
potential-density pairings gives the right-hand side.

\smallskip
\noindent {2.} Monotonicity:
\begin{equation}
(v,\rho), (v',\rho')  \in \zero \; \Rightarrow \;
 \pair{v - v'}{\rho - \rho'} \le 0.
\label{eq:monotonicity}
\end{equation}
If either $(v',\rho)$ or 
$(v,\rho')$ fails to be a ground pair, then the inequality is strict.
This monotonicity inequality\cite{Phelps88,Aubin+Ekeland,Laestadius+18-JCP}
is an immediate specialization of the cross-difference identity,
and generalizes a monotonicity 
previously derived in a DFT context\cite{Gritsenko+Baerends-04,Wagner+13}.

\smallskip
\noindent {3.}
\begin{align}
  (v,\rho)\in\zero & \Rightarrow
                     \nonumber \\
    & \xs({v}',\rho) = E({v}) - E({v}')  + \pair{{v}'-{v}}{\rho}.
\label{eq:v-shift}
\end{align}
This is demonstrated by expanding $\xs({v}',\rho) -  \xs({v},\rho)$ using
the definition (\ref{eq:xs}). 

\smallskip
\noindent {4.}
 \begin{equation}
  (v,\rho)\in\zero_0  \;\Rightarrow\;
 R {v} \in \subdiff_\rho\xs(\tgt{v},\rho).
\label{eq:Rv-in-subdiff}
\end{equation}
Here, $\subdiff_\rho$ denotes the subdifferential with respect to $\rho$ at fixed $v$.
According to the definition of excess energy,
$\subdiff_\rho \xs(\tgt{v},\rho) = \subdiff F(\rho) + \tgt{v}$.
Since $(v,\rho)\in\zero_0$ implies that $-\wh{v} = Rv - \tgt{v} \in \subdiff F(\rho)$,
the conclusion follows.

\section{Modes of approximation}
\label{app:approx}

\begin{appprop}
\label{appprop:energetic-to-residual}
If $\xs(\tgt{v},\rho) < \epsilon$, then $(v',\rho') \in \zero$ for
some $(v',\rho')$ satisfying $\|\tgt{v}-v'\|, \|\rho-\rho'\| < \sqrt{\epsilon}$.
\end{appprop}
\begin{proof}
This is a corollary of the Ekeland variational principle\cite{Ekeland-74}. 
See Cor. I.6.1 of Ref. \onlinecite{ET} or Cor. 5.3.6 of Ref. \onlinecite{Aubin+Ekeland}. 
\end{proof}
\begin{appprop}
\label{appprop:residual-to-energetic}
If $E$ is locally Lipschitz continuous, 
then $(\wh{v},\rho) \in \zero$ and $\wh{v}\in U(\tgt{v})$ 
imply that \hbox{$\xs(\tgt{v},\rho) < (L(\tgt{v})+\|\rho\|) \| Rv\|$},
where $(U(\tgt{v}),L(\tgt{v}))$ are the local Lipschitz data at $\tgt{v}$.
\end{appprop}
\begin{proof}
By definition,
$\xs(\tgt{v},\rho)  - \xs({v},\rho)
= \pair{\tgt{v}-v}{\rho} + E(v) - E(\tgt{v})$.
Since Lipschitz continuity of $E$ means that
$|E(v) - E(\tgt{v})| \le L(\tgt{v}) \|v - \tgt{v}\|$,
the conclusion is immediate.
\end{proof}

\section{Progress}
\label{app:progress}

\subsection{First try}
\begin{appprop}
\label{appprop:progress-bad}
If $({v},\rho), (v+R{v},\rho') \in \zero_0$ and $R{v}\neq 0$, define
  \begin{equation}
    \widetilde{\rho}_\lambda =  (1-\lambda){\rho}_0 + \lambda {\rho}_1, \quad 0\le\lambda.
  \end{equation}
Then, whenever the derivative exists,
\begin{equation}
\frac{d}{d\lambda}\xs(\tgt{v},\widetilde{\rho}_\lambda)\Big|_{\lambda=0} < 0.
\label{eq:progress-1st-try}
\end{equation}
\end{appprop}
\begin{proof}
Apply monotonicity of $\xs_0$ to the two points
$(v,\rho), (v+R v,\rho')\in \zero_0$ (as illustrated in
Fig.~1 of the main text) to obtain
\begin{equation}
\pair{\rho'-\rho}{R{v}} < 0.
\label{eq:mono 2}
\end{equation}
The inequality is strict because $({v}+R v,\rho)\not\in \zero_0$.
For, if both $({v},\rho)$ and $({v}+R v,\rho)$ are in $\zero_0$,
it follows that \hbox{$(\tgt{v},\rho)\not\in \zero$}, contrary to assumption.

Combining (\ref{eq:Rv-in-subdiff}) and (\ref{eq:mono 2}) shows that
$\pair{\rho'-\rho}{\subdiff_\rho\xs(\tgt{v},\rho)}$ contains a negative number. 
Hence, if the derivative exists,
\begin{equation}
\pair{\rho'-\rho}{w}=
\frac{d}{d\lambda}\xs(\tgt{v},\rho+\lambda[\rho'-\rho])\Big|_{\lambda=0}
\end{equation}
for every $w \in {\subdiff_\rho\xs(\tgt{v},\rho)}$.
\end{proof}

\subsection{Progress redux}

\begin{appprop}
\label{appprop:progress-good}
\begin{equation}
\xs(\tgt{v},\rho_\lambda) - \xs(\tgt{v},\rho_0) 
\nonumber
\end{equation}
is equal to
\begin{equation}
 \xs(\wh{v_0},\rho_\lambda)
-\frac{1}{\lambda} 
 \Big[ \xs_0({v_\lambda},\rho_0) + \xs_0({v_0},\rho_\lambda) \Big],
\label{eq:prog-equals}
\end{equation}
and bounded above by either of the following:
\begin{subequations}
\begin{align}
& \lambda^{-1}\pair{{v_\lambda - v_0 }}{\rho_\lambda - \rho_0}  
-\pair{\wh{v_\lambda} - \wh{v_0}}{\rho_\lambda - \rho_0},
\label{eq:bound-a}
\\
&
\pair{(1-\lambda)R v_0 + \slct{\subdiff\Phi(\rho_0)} - \slct{\subdiff\Phi(\rho_\lambda)}}
{\rho_\lambda - \rho_0}.
\label{eq:bound-b}
\end{align}
\end{subequations}
\end{appprop}
\begin{proof}
Apply the identity (\ref{eq:v-shift}) three times, with
$v,\rho,v'$ equal successively to $\wh{v_\lambda},\rho_\lambda,\tgt{v}$, 
$\wh{v_0},\rho_0,\tgt{v}$, and $\wh{v_0},\rho_0,\wh{v_\lambda}$, 
to obtain (every $E$ term occurs once with a plus sign and once with a minus sign)
\begin{align}
\label{eq:step1}
\xs(\tgt{v},\rho_\lambda) - 
\xs(\tgt{v},\rho_0)  & + \xs(\wh{v_\lambda},\rho_0) 
\nonumber \\
= & 
\pair{\tgt{v}-\wh{v_\lambda}}{\rho_\lambda-\rho_0}
\end{align}
Now substitute 
\begin{equation}
\tgt{v} = \wh{v_0} + R{v_0} 
= \wh{v_0} + v_1 - v_0 
= \wh{v_0} + \frac{1}{\lambda}(v_\lambda - v_0)
\nonumber
\end{equation}
into the right-hand side of (\ref{eq:step1}) to find
\begin{align}
  \xs(\tgt{v},\rho_\lambda)
  - &\xs(\tgt{v},\rho_0) 
+ \xs(\wh{v_\lambda},\rho_0) =
\label{eq:step2}
\\
&\frac{1}{\lambda} \pair{v_\lambda - v_0}{\rho_\lambda - \rho_0}
     - \pair{\wh{v_\lambda} - \wh{v_0}}{\rho_\lambda-\rho_0}
     \nonumber
\end{align}
(\ref{eq:prog-equals}) is now obtained by application of the cross-difference
identity (\ref{eq:cross-difference}) to both terms on the RHS of (\ref{eq:step2}).

Dropping the (non-negative) last term on the LHS yields (\ref{eq:bound-a}),
from whence (\ref{eq:bound-b}) follows upon the substitution
\hbox{$\wh{v_\lambda} - \wh{v_0} = {v_\lambda} - {v_0} +
 \slct{\subdiff\Phi(\rho_0)} - \slct{\subdiff\Phi(\rho_\lambda)}$}.
\end{proof}

\section{Analyticity in standard interpretation}
\label{sec:app-Lieb}

The Lieb interpretation\cite{Lieb83} of our abstract framework is the standard
mathematical theory of quantum mechanical ground-state DFT for a fixed number $N$ of
identical particles. In the Lieb theory, the space $B$ of densities is the real Banach space
$B \defeq L^{3}(\Reals^3)\cap L^1(\Reals^3)$. The norm of $f\in B$ is
the sum $\|f\|_B = \|f\|_{L^1} + \|f\|_{L^3}$ of its $L^1$ and $L^3$ norms.
The dual space is $B' = L^{3/2}(\Reals^3)+L^\infty(\Reals^3)$ consisting of real functions
which can be written as a sum of functions in $L^{3/2}$ and $L^\infty$, and the norm is
\begin{equation}
  \label{eq:B'-norm}
\|v\|_{B'} = \inf \setof{\|v'\|_{L^{3/2}} + \|v''\|_\infty}{v'+v'' = v}.  
\end{equation}
Normally, one defines the norm on a dual space $X'$ as
$\|\lambda \|_{X'} = \sup_{\|x\|=1} \pair{\lambda}{x}$.
The definition (\ref{eq:B'-norm}) does not satisfy this, but it is equivalent,
in the sense that the two are mutually bounded, hence define the same topology
on $B'$. 

Of course, densities and potentials ought to be real. However, we will have
use for complexified versions in the following, which will be indicated
by a subscript, as $B_\Cmplxs$ and $\VC$.
The set of external potentials in $B'$ such that the corresponding
reference system Hamiltonian has an isolated ground state eigenvalue
and the ground state manifold comprises a single spin multiplet will be
denoted $\Vzero$. This is a very important subset; both main results
are concerned only with it.
If $v\in \Vzero$, then there is a unique ground density;
it will be denoted $\rho[v]$, square brackets being used simply because
the normal argument of a density is position.

The essential conclusions of this Section are as follows.
\begin{appprop}
\label{appprop:holomorphy-Lieb}
${\mathscr V}_0$ is open, and $\rho[v]$, $F(\rho[v])$, and 
$\xs(\tgt{v},\rho[v])$ are analytic functions of $v$ on ${\mathscr V}_0$.
\end{appprop}
\begin{appcor}
  \label{cor:progress-Lieb}
Suppose $v_0 \in {\mathscr V}_0$, and define $v_\lambda \defeq v_0 + \lambda Rv_0$. 
There is $\epsilon>0$ such that, for $|\lambda|<\epsilon$,
$v_\lambda \in {\mathscr V}_0$, and therefore $\rho_\lambda \defeq \rho[v_\lambda]$ is unambiguous.
\newline
Then, either $\frac{d}{d\lambda}\xs(\tgt{v},\rho_\lambda) < 0$, or 
$\rho_\lambda = \rho_0$ for $|\lambda|<\epsilon$.
The latter happens only if the ground state(s) are eigenstates of $Rv_0$.
\end{appcor}

These will be proved in Section \ref{sec:proofs} after reviewing/collecting some
tools. This material is more technically dense than the previous Sections,
albeit of a sort which may be more familiar.
Prop. \ref{appprop:holomorphy-Lieb} depends on none of the preceding, while
Cor. \ref{cor:progress-Lieb} depends on Props. 
\ref{appprop:progress-good} and \ref{appprop:holomorphy-Lieb}.

Essentially the problem is that we deal with a family of unbounded operators,
which do not even have a common domain, and the chosen solution is to introduce
appropriate auxiliary spaces so that everything is expressed in terms of bounded
operators. Section \ref{sec:rigging} reviews the technique known variously under

\subsection{Kinetic energy Hilbert rigging}
\label{sec:rigging}

We will use the method of rigged Hilbert spaces (``scale of spaces'', ''Sobolev tower'',
``Gelfand triple''),
and give a brief account tailored to the immediate needs. For systematic expositions, see 
Refs. \onlinecite{Kato,Reed+Simon,Simon-forms,deOliveira,Schmudgen}.

  With an appropriate choice of units, the operator representing kinetic energy
  is the $3N$-dimensional Laplacian $-\Delta$, acting on the Hilbert space
  $\Hilb_0 \defeq L^2(\Reals^{3N})$ with the usual
  inner product $\inpr{{\psi}}{{\phi}}_0 = \int \cc{\psi(x)}\phi(x)\, d^{3N}x$.
  The basic idea now is to work with a triplet
  $\Hilb_+ \subset \Hilb_0 \subset \Hilb_-$
of Hilbert spaces,
where $\Hilb_+$ consists of ``smooth'' vectors which will be common sesquilinear
form domain of all of our Hamiltonians, while $\Hilb_-$ consists of ``generalized''
vectors and is identified with the dual space of $\Hilb_+$ with respect to the inner
product $\inpr{\phantom{\psi}}{\phantom{\phi}}_0$.

Following physics custom, momentum representation is indicated by argument ($p$) rather
than a notation for Fourier transformation, so the kinetic energy acts as
  \begin{equation}
{(-\Delta \psi)}(p) = |p|^2 {\psi}(p).
\end{equation}
This does not define an element of $\Hilb_0$ unless
$\int |p|^4 |{\psi}(p)|^2\, d^{3N}p$ is finite, which condition delimits the
operator domain of $-\Delta$. On the other hand, the sesquilinear form (conjugate linear in
first argument, linear in second) 
  \begin{equation}
    \nonumber
    \Dbraket{\psi}{-\Delta}{\phi}_0
    = \int \nabla\cc{\psi}\cdot\nabla{\phi}\, d^{3N}x    
    = \int |p|^2 \cc{{\psi}(p)}{\phi(p)}\, d^{3N}p,    
  \end{equation}
  is well-defined for $\psi$ and $\phi$ in the larger subspace $Q(-\Delta)$ of
  $\Hilb_0$ consisting of wavefunctions satisfying merely
  $\int |p|^2 |{\psi}(p)|^2\, d^{3N}p < \infty$.
Equipping $Q(-\Delta)$ with the inner product  
  \begin{equation}
    \inpr{{\psi}}{{\phi}}_+ \defeq
    \inpr{{\psi}}{{\phi}}_0 + \Dbraket{\psi}{-\Delta}{\phi}_0 =
 \Dbraket{\psi}{1-\Delta}{\phi}_0,
  \end{equation}
  it becomes a Hilbert space, denoted $\Hilb_+$, with the norm $\|\phantom{\psi}\|_+$.
We are making an idiosyncratic use of Dirac notation here:
  $\inpr{\psi}{A\phi}$ implies that $A\phi$ is actually a vector in the Hilbert
  space, whereas $\Dbraket{\psi}{A}{\phi}$ is a sesquilinear form.

  Elements of $\Hilb_+$ are also elements of $\Hilb_0$, so there is a natural injection
\begin{equation}
\label{eq:iota+}
{\iota_+}\colon{\Hilb_+}\hookrightarrow {\Hilb_0},
\end{equation}
which is bounded: $\|\iota_+ \psi\|_0 \le \|\psi\|_+$.
Since $\iota_+$ changes the way we regard the wavefunction $\psi$, 
but not $\psi$ qua function, it will usually be omitted unless confusion would result.

Now we need to consider the dual space of $\Hilb_+$, i.e., the space of continuous linear
functionals. The Riesz representation theorem teaches that we can identify that dual with
$\Hilb_+$, relative to $\inpr{\phantom{\phi}}{\phantom{\phi}}_+$.
That is, as $\phi$ ranges over functions such that $\sqrt{1+|p|^2}{\phi}(p)$ is
square integrable, 
\begin{equation}
\psi \mapsto  \inpr{\phi}{\psi}_+ = \int (1+|p|^2) \cc{{\phi}(p)}  {\psi}(p)\, d^{3N} p
\end{equation}
ranges over all continuous linear functionals on $\Hilb_+$.
On the other hand, as $\phi$ ranges over $\Hilb_+$,
$({J_+^-\phi})(p) \defeq (1+|p|^2){\phi}(p)$
ranges over functions $\Phi(p)$ such that $(1+|p|^2)^{-1/2} \Phi(p)$ is square-integrable.
This motivates defining yet another inner product,
\begin{equation}
  \inpr{\phi}{\psi}_- = \int  \cc{{\phi}(p)}  {\psi}(p)\,  \frac{d^{3N} p}{1+|p|^2},
\end{equation}
and Hilbert space $\Hilb_-$ of functions ${\psi}(p)$ such that the associated norm
$\|\psi\|_- < \infty$ is finite.
Beware: for $\psi\in \Hilb_-$, even though $\psi(p)$ is a {\em function},
$\psi(x)$ might be just a distribution.
One need look no further than the familiar $\delta$ ``function'' to see an example of this phenomenon.
(However, $\delta$ is too singular to be in $\Hilb_-$ unless the spatial dimension is
less than $4/N$.)
Just as $\Hilb_+$ is continuously embedded into $\Hilb_0$ via $\iota_+$,
$\Hilb_0$ is continuously embedded into $\Hilb_-$, and we call this mapping $\iota_0$.
In addition, the map $J_+^-$ introduced above is a unitary mapping between $\Hilb_+$
and $\Hilb_-$ with inverse $J_{-}^{+}$. Summing up, for $\phi,\psi \in \Hilb_+$,
\begin{equation}
  \inpr{J_+^-\phi}{J_+^-\psi}_-
  =  \inpr{\phi}{\psi}_+
= \inpr{J_+^-\phi}{\psi}_0.
\end{equation}

Now we add potentials to the picture.
$v$ in $B'$, a priori merely a function on $\Reals^3$, is turned into
a proper one-body potential as $\Gamma_{\text{ext}}^0 v(x_1,\ldots,x_N) = \sum_n v(x_n)$.
Similarly, $\Gamma_{\mathrm{int}}^0$ turns it into a two-body interaction.
These potentials can be bounded as (subscript $*$ stands for $\mathrm{ext}$ or $\mathrm{int}$)
\begin{equation}
  \label{eq:form-bound}
  |\Dbraket{\psi}{\Gamma_{*}^0v}{\psi}_0| \le a \|\psi\|_0^2
  + b\|v\|_{B'} \|{\psi}\|_+^2,
\end{equation}
for all $\phi$ and $\psi$ in $\Hilb_+$, where
(i) for any $v\in B'$, $b>0$ can be taken as small as desired, at the cost of making
$a$ large, and (ii) there is $b_0$ such that $(a,b)=(0,b_0)$ works uniformly for all $B'$.
For these properties of the bound, see Lemma VI-4.8b of Kato's treatise\cite{Kato}.

This has the following consequences. First, there are {\em bounded}
operators $\Arr{B'}{\Gamma_*^+}{B(\Hilb_+}$ such that ($\psi,\phi\in\Hilb_+$)
\begin{equation}
\Dbraket{\phi}{\Gamma_{*}^0w}{\psi}_0  =  \inpr{\phi}{(\Gamma_{*}^+w) \psi}_+.
\end{equation}
Secondly, for fixed $v\in B'$, there is $m(v)$ such that the norm
\begin{equation}
\label{eq:re-norm}
\|{\psi}\|_v^2 = 
m\|{\psi}\|_0^2
+ \|{\psi}\|_+^2
+ \Dbraket{\psi}{\Gamma_{\text{ext}}^0v}{\psi}_0
\ge \|{\psi}\|_0^2
\end{equation}
is equivalent to the norm $\|\phantom{\psi}\|_+$ on $Q(-\Delta)$.
Thus, the triplets of spaces generated by $-\Delta + \Gamma_{\mathrm{ext}}v$
and $-\Delta$ are equivalent. Except for the explicit momentum-space
representation, everything said prior to this point holds equally for either.
Therefore, in the following, when we will have some fixed $v\in B'$ in mind,
$\|\phantom{\psi}\|_+$ will really mean $\|\phantom{\psi}\|_v$ as given
in (\ref{eq:re-norm}) and $\Gamma_*^+$ will be defined relative to it.


%

\subsection{Analyticity and holomorphy}
\label{sec:analyticity}

We recall some important basic notions of differential calculus in
Banach spaces. Textbook treatments can be found in many places,
such as Refs. \onlinecite{Lang-Real_analysis,AMR,AMP,Mujica}. See Mujica's book
for holomorphy.
Let $X$ and $Y$ be Banach spaces, $U$ an open subset of $X$ and
$\Arr{U}{f}{Y}$ a function.
The Fr\'echet derivative of $f$ at $a$ is the unique bounded linear map
$\Arr{X}{Df(a)}{Y}$ satisfying
\begin{equation}
  f(a+x) = f(a) + Df(a) x + o(\|x\|),
\end{equation}
assuming such exists.
$f$ is said to be {\em differentiable} on $U$ if $Df(a)$ exists for every $a\in U$.
In that case, $Df$ is a function from $U$ into the Banach space $L(X;Y)$
and with sufficient regularity, the construction can be repeated to
obtain the second derivative $\Arr{U}{D^2f}{L(X;L(X;Y))}$.
The codomain here is naturally isometric to the space $L(X\times X;Y)$
of continuous bilinear mappings from $X$ into $Y$, and that is the preferred
way to view it, since $D^2f$ is symmetric in its arguments.
Higher derivatives $\Arr{U}{D^nf}{L(\overbrace{X\times\cdots\times X}^{n};Y)}$
are defined by continuing the pattern.
If derivatives of all orders exist at $a$, and
\begin{equation}
  f(a+x) = \sum_{n=0}^\infty D^nf(a)(\overbrace{x,\cdots,x}^{n}).
  \label{eq:analytic}
\end{equation}
uniformly for $\|x\|$ small enough, then $f$ is {\em analytic} at $a$, and
it is analytic on $U$ if analytic at each point of $U$.

So far, no distinction has been made between $\Reals$ or $\Cmplxs$ as the scalar field.
Suppose that $X$ and $Y$ are complex spaces. Even so, $f$ might only be
$\Reals$-differentiable, $\Reals$-analytic, etc.
However, if $f$ is assumed to be merely $\Cmplxs$-differentiable on $U$, then
$\Cmplxs$-analyticity (holomorphy) follows automatically. In fact, 
sufficient conditions can be reduced to ones involving one-dimensional
domain and range spaces as follows.
$f$ is said to be {\em weakly G-holomorphic} on $U$ if
the map $\zeta \mapsto \pair{\lambda}{f(a+\zeta x)} \colon \Cmplxs\rightarrow\Cmplxs$ is
$\Cmplxs$-differentiable at zero for every $a\in U$, $x\in X$ and $\lambda\in Y'$.
If $f$ is $G$-holomorphic on $U$ and locally bounded, then it is holomorphic
[i.e., in the sense of (\ref{eq:analytic})].

\subsection{Proofs}
\label{sec:proofs}

The proof of Prop. \ref{appprop:holomorphy-Lieb} and Cor. \ref{cor:progress-Lieb}
is organized into several numbered steps. Only the spinless case is considered
initially. Spin is incorporated in the final Step.

\begin{figure}
  \centering
\includegraphics[width=60mm]{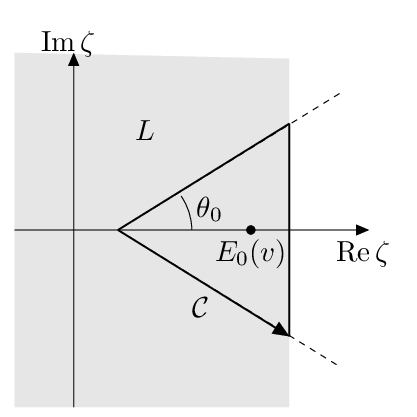}  
\caption{Geometry in the complex plane of the spectral parameter $\zeta$ relevant to
  the construction of holomorphic spectral projectors. $L$ is the shaded left half-plane and
  for $u\in\UU$, the spectrum of $H(v+u)$ in $L$ is actually within ${\mathcal C}$.
}
  \label{fig:spectral}
\end{figure}

{\em Fix} $v\in {\mathcal V}_0$. By definition, $H(v) \defeq -\Delta + \Gamma^0_{\mathrm{ext}}v$
has nondegenerate ground state eigenvalue $E_0(v)$,
and for some $\epsilon$, the part of $\Spec H_v$ in the left half-plane
$L \defeq \{\re \zeta \le E_0(v)+\epsilon\}$ consists only of that eigenvalue.
See Fig. \ref{fig:spectral} for an illustration of this and later points.
For convenience, take $E_0(v) > 0$ by adding a constant to $v$ if necessary;
then, we can choose simply [see (\ref{eq:re-norm})]
\begin{equation}
\inpr{\psi}{\phi}_+ = \Dbraket{\psi}{H(v)}{\phi}_0.  
\end{equation}
We are interested in perturbations of $v$ by complex pontentials $u$ in some
neighborhood $\UU$ of zero in $\VC$. Several conditions imposed potentially
limit the size of $\UU$ which will be flagged with the annotation ``[shrink]''; 
the reader may imagine $\UU$ being implicitly shrunk at such points of the discussion.
For $Y$ a Banach space, $\Hol(\UU;Y)$ denotes the set of holomorphic functions
from $\UU$ to $Y$.

\begin{enumerate}
\item[1.] 
For $u\in\UU$, $H(v+u)$ has a nondegenerate ground state vector $\psi[v+u]$. 
\hbox{$(u\mapsto \psi[v+u]) \in \Hol(\UU,\Hilb_0)$} and 
\hbox{$(u\mapsto E_0(v+u)) \in \Hol(\UU,\Cmplxs)$}.
\end{enumerate}

In its essence, this is a well-established form of perturbation theory,
see \S VII.4 of Kato's treatise\cite{Kato}.
A major difference is that our family of perturbations $u$ is much larger than
the traditional family $zu$ for {\em fixed} $u$ and complex $z$, but this causes
surprisingly little difficulty. Our relatively self-contained exposition here
has stylistic differences.

The estimate (\ref{eq:form-bound}) shows that
the spectrum of $H(v+u)$ for $u$ in $\UU$ [shrink],
is contained in a right-facing wedge in the complex plane,
as illustrated in Fig. \ref{fig:spectral}.
The basic plan is to use the formula
\begin{equation}
  \label{eq:P(u)}
P(u) = \oint_{\mathcal C} [\zeta - H(v+u)]^{-1} \frac{d\zeta}{2\pi i},
\end{equation}
to construct a (non-orthogonal, in general) spectral projection $P(u)$ for $H({v+u})$,
where the contour ${\mathcal C}$ in the complex $\zeta$-plane is shown in
Fig. \ref{fig:spectral}. For $u\in \UU$, $P(u)$ is
well-defined and holomorphic in $u$, actually corresponds to a
spectral projection for the entire region $L$ (see earlier remarks on where
the spectrum is),
and that spectrum consists of a single nondegenerate eigenvalue
because range dimension is a continuous function on projectors
(see \S XII.2 of Reed\& Simon\cite{Reed+Simon} or \S I.4.6 of Kato\cite{Kato}).
Given $P(u)$, we then obtain a ground state vector by
\begin{equation}
  \label{eq:psi[v+u]}
\psi[v+u] = \frac{P(u)\psi[v]}{\inpr{\psi[v]}{P(u)\psi[v]}^{1/2}},  
\end{equation}
which is normalized for real $u$.

The problem therefore reduces to showing holomorphy
of the resolvent operator $[\zeta - H(v+u)]^{-1}$ for $\zeta$
along ${\mathcal C}$ and $u\in \UU$.
Our solution involves representing the {\em graphs} of $H(v+u)$ as bounded operators
on a common domain and manipulating those representations.
Recall that for $\psi$ and $\phi$ in $\Hilb_+$,
$\Dbraket{\psi}{-\Delta + \Gamma_{\mathrm{ext}}^0v + \Gamma_{\mathrm{ext}}^0u}{\phi}_0
= \Dbraket{\psi}{1 + \Gamma_{\mathrm{ext}}^+u}{\phi}_+$, and that
$\Gamma_{\mathrm{ext}}^+$ is a bounded operator, so that for small enough
$\|u\|_{\mathscr V}$, ${1 + \Gamma_{\mathrm{ext}}^+u}$ is invertible.
Thus, we can construct the following chain of linear operators
\begin{equation}
  \label{eq:chain}
T(u)\colon \Hilb_{0} \xhookrightarrow{\iota_0}  \Hilb_{-} 
\xrightarrow{J_-^+} \Hilb_{+} 
\xrightarrow{(1 + \Gamma_{\mathrm{ext}}^+u)^{-1}}\Hilb_+
\xhookrightarrow{\iota_+}  \Hilb_{0}.
\end{equation}
The composite,
$T(u) \defeq {\iota_+}\circ(1 + \Gamma_{\mathrm{ext}}^+u)^{-1}\circ{J_-^+}\circ{\iota_0}$,
is $H(u+v)^{-1}$.
To verify this, take
$\phi\in\Hilb_0$ and $\psi\in\Hilb_+$. Then,
\begin{align}
  \inpr{\psi}{H(v+u)T(u)\phi}_0
  &=
    \inpr{\psi}{H(v+u)(1 + \Gamma_{\mathrm{ext}}^+u)^{-1}{J_-^+}\iota_0\phi}_0
    \nonumber \\
  &=
    \inpr{\psi}{(1+\Gamma_{\mathrm{ext}}^+u) (1 + \Gamma_{\mathrm{ext}}^+u)^{-1}{J_-^+}\iota_0\phi}_+
    \nonumber \\
  &=
    \inpr{\psi}{{J_-^+}\iota_0\phi}_+
    \nonumber \\
  &=
    \inpr{\psi}{\phi}_0
    \nonumber.
\end{align}
By density of $\Hilb_+$ in $\Hilb_0$, this shows that $H(v+u)T(u) = 1$.
$T(u)H(v+u) = 1$ is shown similarly.
Since composition, and inversion where possible, preserve holomorphy,
$T(u)$ is holomorphic.

The conclusion of the preceding can be rephrased as: the linear map
\begin{equation}
\Arr{\Hilb_0}{(1,T(u))}{\Hilb_0\times\Hilb_0}
\end{equation}
is a holomorphic (in $u$) parametrization of the graph of $H(v+u)^{-1}$.
The reason for this silly-looking rephrasing is to obtain the
resolvent by manipulating this graph.
First, $(T(u),1)$ parametrizes the graph of $H(v+u)$, so
$(1-\zeta T(u),T(u))$ parametrizes the graph of the resolvent
$[H(v+u)-\zeta]^{-1}$ when $\zeta$ is in the resolvent set $\mathrm{Res}\, H(v+u)$.
(Otherwise, it's not a {\em graph} at all.)
And, therefore, $T(u)[1-\zeta T(u)]^{-1}$ simply {\em is} the resolvent,
whenever $1-\zeta T(u)$ is invertible.
Suppose, then, that $\zeta_0 \in \mathrm{Res}\, H(v)$.
In that case, $1-\zeta_0 T(0)$ is invertible, and, therefore,
$1-\zeta T(u)$ is invertible for $(\zeta,u)$ in some neighborhood of $(\zeta_0,0)$.
By a compactness argument, the contour ${\mathcal C}$,
as in Fig. \ref{fig:spectral}, is in the resolvent set for every $u$ in $\UU$ [shrink].
For such $u$,
$T(u)[1-\zeta T(u)]^{-1}$ is now established as a plainly holomorphic expression
for the resolvent on ${\mathcal C}$, and therefore $P(u)$ in (\ref{eq:P(u)})
is also holomorphic.
Then, since $T(u) = H(v+u)^{-1}$ is holomorphic, so is
\begin{equation}
E_0(v+u)^{-1}
  = \frac{ \inpr{\psi[v]}{T(u)P(u)\psi[v]} }{ \inpr{\psi[v]}{P(u)\psi[v]}  },
\end{equation}
as well as its inverse. 

\begin{enumerate}
\item[2.] 
\begin{equation}
(  u \mapsto \psi[v+u] ) \in \Hol(\UU;\Hilb_+)
\end{equation}
\end{enumerate}

This makes no sense unless $\psi[v+u]$ is actually in $\Hilb_+$,
but that follows immediately from (\ref{eq:chain}). To finish, we use the
equivalence of strong and weak holomorphy.
We need to show that
$\inpr{\phi}{\psi[v+u]}_+$ is holomorphic for every $\phi\in\Hilb_+$.
Because $1+\Gamma_{\mathrm{ext}}^+u$ has a holomorphic inverse, 
that is equivalent to holomorphy of $\inpr{\phi}{(1+\Gamma_{\mathrm{ext}}^+u)\psi[v+u]}_+$.
However,
\begin{align}
\inpr{\phi}{(1+\Gamma_{\mathrm{ext}}^+u)\psi[v+u]}_+
  &= \Dbraket{\phi}{H(v+u)}{\psi[v+u]}_0
    \nonumber \\
  &= E(v+u) \inpr{\phi}{\psi[v+u]}_0,
    \nonumber
\end{align}
and the final expression, as a product of holomorphic functions, is holomorphic.

\begin{enumerate}
\item[3.] 
\begin{equation}
( u \mapsto \rho[v+u] ) \in \Hol(\UU;B_\Cmplxs).
\end{equation}
\end{enumerate}
This requires defining $\rho[v+u]$. For non-real $u$, we cannot simply
substitute $\psi[v+u]$ into the formula
$N\int |\psi(x,x_2,\ldots,x_N)|^2  \, dx_2\cdots dx_N$ for density.
That could not possibly be holomorphic because composition with a continuous
antilinear map interchanges holomorpy and antiholomorphy.
Instead of $\cc{\psi[v+u]}$, we need $\cc{\psi[v+\cc{u}]}$.
For real $u$, that changes nothing.

Use use equivalence of weak and strong holomorphy, again. 
It suffices to show that for every $w\in \VC$, $\pair{w}{\rho[v+u]}$ is holomorphic in $u$:
\begin{align}
  \pair{w}{\rho[v+u]}
    &  = \Dbraket{\psi[v+\cc{u}]}{\Gamma_{\mathrm{ext}}^0 w}{\psi[v+\cc{u}]}_0
      \nonumber \\
& = \inpr{\psi[v+\cc{u}]}{(\Gamma_{\mathrm{ext}}^+ w) \psi[v+\cc{u}]}_+,
  \end{align}
  and the final expression is holomorphic by Step {2}.

\begin{enumerate}
\item[4.] 
\begin{equation}
( u \mapsto F(\rho[v+u]) ) \in \Hol(\UU;\Cmplxs).
\end{equation}
\end{enumerate}
We have
\begin{equation}
F_0(\rho[v+u])) = E_0(v+u) - \pair{v+u}{\rho[v+u]},
\end{equation}
and both terms on the right-hand side have already been shown holomorphic.
If $w\in\V$ is the interaction potential, then
$\Dbraket{\psi[v+\cc{u}]}{\Gamma_{\mathrm{ext}}^0 w}{\psi[v+\cc{u}]}_0$ is
shown holomorphic by a calculation like that in Step {3}.

\begin{enumerate}
\item[5.] 
\begin{equation}
( u \mapsto \xs(\tgt{v},\rho[v+u]) ) \in \Hol(\UU;\Cmplxs).
\end{equation}
\end{enumerate}
This now follows trivially from Steps {3} and {5}.

The proof of Prop. \ref{appprop:holomorphy-Lieb} now requires just a
little cleanup. $v$ was arbitrary in ${\mathscr V}_0$, and
$\UU \cap B'$ is a neighborhood of $v$ on which $\rho[v]$, $F(\rho[v])$,
$\xs(\tgt{v},\rho[v])$ are real analytic, so ${\mathscr V}_0$ is open
and those functions are analytic on all of it.

\begin{enumerate}
\item[6.] 
Define $v_\lambda \defeq v + \lambda Rv$ 
and $\rho_\lambda \defeq \rho(v_\lambda)$.
Either $\frac{d}{d\lambda}\xs(\tgt{v},\rho_\lambda)\Big|_0 < 0$, or $\rho_\lambda = \rho_0$.
The latter can happen only if the ground state is an eigenstate of $Rv$.
\end{enumerate}

Prop. \ref{appprop:progress-good} enters the discussion at this point.
According to it,
\begin{align}
  \label{eq:6}
\xs(\tgt{v},\rho_\lambda) - \xs(\tgt{v},\rho_0) 
 = & \xs(\wh{v_0},\rho_\lambda)
  \\
& -\frac{1}{\lambda} 
 \Big[ \xs_0({v_\lambda},\rho_0) + \xs_0({v_0},\rho_\lambda) \Big].
\nonumber
  \end{align}
  The excess energies here can be expressed as quantum mechanical quadratic
  forms. For instance,
  \begin{equation}
\xs(\wh{v_0},\rho_\lambda) =
\Dbraket{\psi_\lambda}{H(\wh{v_0})+\Gamma_{\mathrm{int}}^0 w-E(\wh{v_0})}{\psi_\lambda}_0.
\end{equation}
Since $\psi_\lambda$ is analytic as a vector in $\Hilb_+$, such expressions
can safely be manipulated in an apparently naive way.
So,
\begin{align}
  \xs(\wh{v_0},\rho_\lambda)
  &=
\lambda 2 \re \Dbraket{\dot{\psi}_0}{H(\wh{v_0})+\Gamma_{\mathrm{int}}^0 w-E(\wh{v_0})}{{\psi}_0}_0
    + \bigO(\lambda^2)
    \nonumber \\
= \bigO(\lambda^2),
   \nonumber
\end{align}
where over-dot denotes differentiation with respect to $\lambda$.
Similarly,
\begin{equation}
\xs_0({v_0},\rho_\lambda) = \lambda^2 \|\dot{\psi}_0\|_+^2 + \bigO(\lambda^3),
\end{equation}
and
\begin{align}
  \xs_0({v_\lambda},\rho_0)
  &= 
    \Dbraket{\psi_0}{H({v_0})+\lambda\Gamma_{\mathrm{ext}}^0 Rv_0 - E(v_\lambda)}{\psi_0}_0
    \nonumber \\
&= \lambda \Dbraket{\psi_0}{\Gamma_{\mathrm{ext}}^0 Rv_0 - \dot{E}(v_0)}{\psi_0}_0
                   +\bigO(\lambda^2).
\label{eq:excess-v0}
\end{align}

Now, since the left-hand side of (\ref{eq:6}) is $\bigO(\lambda)$, the
$\xs_0({v_\lambda},\rho_\lambda)$ must vanish to $\bigO(\lambda)$. 
That recovers the usual first-order formula for energy shift,
$\dot{E}(v_0) =  \Dbraket{\psi_0}{\Gamma_{\mathrm{ext}}^0 Rv_0}{\psi_0}_0$,
and since $\xs_0({v_\lambda},\rho_\lambda) \ge 0$, the $\bigO(\lambda^2)$ term in 
(\ref{eq:excess-v0}) must be non-negative.
Hence,
\begin{equation}
\xs(\tgt{v},\rho_\lambda) - \xs(\tgt{v},\rho_0) \le 
-\lambda \|\dot{\psi}_0\|_+^2 + \bigO(\lambda^2),
\end{equation}
and the derivative of (\ref{eq:6}) is negative, unless
$\dot{\psi}_0=0$.
To see what the implications of that would be, consider
\begin{equation}
\Dbraket{\phi}{H({v_\lambda})-E({v_\lambda})}{{\psi}_\lambda}_0=0,
\end{equation}
for $\phi\in\Hilb_+$. Differentiating this and assuming $\dot{\psi}=0$,
\begin{equation}
\Dbraket{\phi}{\Gamma_{\mathrm{ext}}^0Rv_0-\dot{E}({v_0})}{{\psi}_0}_0=0.
\end{equation}
$\Hilb_+$ being dense in $\Hilb_0$, this implies that $\psi_0$ is an eigenvector of
$\Gamma_{\mathrm{ext}}^0Rv_0$, and therefore of $H(v_\lambda)$ for all $\lambda$.
The ground state being unique by assumption, $\psi_0$ is it.

\begin{enumerate}
\item[7.] Add spin.
\end{enumerate}

Since spin rotation commutes with the kinetic energy and all the potentials under
consideration, the Hilbert space decomposes into a direct sum of spin sectors.
On each one the picture of the preceding discussion holds, with a uniform degeneracy.
One need only say, perhaps, that $\UU$ needs to be shrunk a bit more to ensure than
no spin sector other than that containing the ground state obtains spectrum inside
the contour $\mathcal C$ of Fig. \ref{fig:spectral}.

\bibliographystyle{apsrev4-1}
\bibliography{../Bib/dft,../Bib/fa,../Bib/convex}


\end{document}